# Indicators as judgment devices: Citizen bibliometrics in biomedicine and economics

Björn Hammarfelt [1,2] & Alexander D. Rushforth[2]

*The number of publications has been a fundamental merit in the competition for academic positions since the late 18th century. Today, the simple counting of publications has been supplemented with a whole range of bibliometric measures, which supposedly not only measures the volume of research but also its impact. In this study, we investigate how bibliometrics are used for evaluating the impact and quality of publications in two specific settings: biomedicine and economics. Our study exposes the extent and type of metrics used in external evaluations of candidates for academic positions at Swedish universities. Moreover, we show how different bibliometric indicators, both explicitly and implicitly, are employed to value and rank candidates. Our findings contribute to a further understanding of bibliometric indicators as 'judgment devices' that are employed in evaluating individuals and their published works within specific fields. We also show how 'expertise' in using bibliometrics for evaluative purposes is negotiated at the interface between domain knowledge and skills in using indicators. In line with these results we propose that the use of metrics in this context is best described as a form of 'citizen bibliometrics' – an underspecified term which we build upon in the paper.*

## 1. Introduction

Since the 1970s much of the promise of evaluative bibliometrics (Narin 1976) has been premised on the notion of tempering the subjective and cognitive biases of peer review, so much so that it has often been imagined as an alternative mode of evaluating. In practice however, bibliometrics tends to supplement expert decision-making rather than supplant it (Moed 2007; van Raan 1996). Indeed calls to use (advanced) bibliometrics as part of 'informed peer review' processes has been posited as a means of mitigating the weaknesses of both approaches (Butler 2007). At the same time, there are often assumptions made that simple output indicators like Journal Impact Factor (JIF), h-index, and journal ranking lists are commonly used in decision–making contexts. Despite such assumptions, to date few have responded to earlier calls by Woolgar (1991) to study actual uses of indicators in peer review and other decision–making contexts.

[1]Swedish School of Library and Information Science, SSLIS, University of Borås, SE-501 90 Borås, email: bjorn.hammarfelt@hb.se(Corresponding author)

[2]CWTS, Leiden University, 2333 AL Leiden, The Netherlands, e-mail: a.d.rushforth.cwts.leidenuniv.nl

Whilst some attention has been directed towards researchers' attitudes towards bibliometrics (Aksnes and Rip 2009; Buela-Casal and Zych 2012) fewer still have studied actual uses of bibliometrics and its consequences for knowledge production (Rushforth and de Rijcke 2015).

Studies regarding the formalized uses of metrics in research assessments are more common, and a literature looking at practices and effects is gradually emerging (de Rijcke, Wouters, Rushforth, Franssen, and Hammarfelt 2015). While acknowledging the importance of these approaches we suggest that metrics might have even more profound influence on the micro–level of individuals and smaller groups. For this reason it is important to engage with the uses of metrics in high–stake contexts, where employing bibliometric indicators might have major consequences for the individual researcher.

Our main focus in this paper is the uses of metrics in forming judgments of applicants for academic positions. More specifically we investigate how bibliometric indicators are used for ranking candidates in two specific settings: biomedicine and economics. Based on qualitative analysis of written assessment reports of applicants, a first set of issues addressed in our study concerns questions such as: To what extent are bibliometric measures used to evaluate candidates for academic positions? In what ways are these measures used? And how are different indicators compared, negotiated and discussed?

Our findings hope to elucidate the extent and type of bibliometrics used for evaluation purposes, and in doing so open-up understanding of how individuals are evaluated. Our selection of fields is motivated by an ambition to study disciplines that both draw on metrics, but which differ in their social and intellectual structure. Building on the works of Whitley (Whitley, 2000; Whitley and Gläser 2008) we infer that differences in the organization of research fields is likely to have direct consequences for the forming of evaluation practices. The degree of dependency, heterogeneity in research practices and publication strategies, as well as the agreement on research goals and methods are some of the factors that are likely to influence the assessment of research. The widespread presence of bibliometrics in biomedicine, and economics, has been widely reported in the scientometric research literature, mostly by way of technical discussions about measures used to evaluate outputs (Graber, Launov, and Wälde 2008; Haucap and Muck 2015). The broad coverage of biomedical literature in *Web of Science*, and later *Scopus*, together with the sheer size of the field has contributed to a frequent use of performance indicators in the field of biomedicine (De Bellis 2009; Van Eck, Waltman, van Raan, Klautz, and Peul 2013;). The intellectual organization of economics differs when compared with other social sciences (Whitley 2000; Fourcade, Ollion & Algan 2015), for instance publication patterns and citation patterns in economics allow for the use of citation databases on a much more widespread scale than other social science disciplines (Hicks 2004). Yet, although several studies points to the influence of metrics in these fields we find little research on how indicators are used to assess and rank individual researchers.

Despite a proliferation in metrics across all kinds of evaluation contexts (Beer 2016), notions of quality remain multi-dimensional (Musselin 2009). In this study we relate the complexity of ranking candidates to the heterogeneity of entities being evaluated; all candidates are unique and no single criteria can be used to make judgments. The process of evaluating, and ranking candidates, could be compared to the valuation of what Karpik (2010) calls 'singularities': unique products that cannot easily be compared – art, literary works, medical doctors etc. – or valued on a market. Their valuation is therefore dependent on 'judgment devices' to facilitate a uniform ranking of items. Consequently, we propose that bibliometric indicators can be viewed as similar devices, which are used as aids for making decisions when hiring academic personnel. As such we show how expertise in these documents is not so much outsourced or eroded by bibliometrics, but rather gets re–defined through them in quite varied forms. Based on this analysis, we hope to contribute to emerging debates on what tentatively might be called 'citizen bibliometrics' (c.f. Wouters et al. 2015; Leydesdorff, Wouters and Bornman, in press). 'Citizen bibliometrics' is so far an underspecified concept, but in our view it has both normative and descriptive implications. Citizen bibliometrics is normative in the sense that it relates to being a citizen, to being part of a collective that has certain rights but also duties. Building on the notion of 'academic citizenship' (MacFarlane 2007), we could therefore infer that citizen bibliometrics suggest that a certain responsibility, or even care, for one's discipline is implicated in the concept. Importantly 'citizenship' here relates less to the morality of individuals and more to the ethics of a collective. These normative aspects of 'citizen bibliometrics' played a role in our decision to use the term, but primarily the concept serves as analytical perspective which allows us to examine and question the oft-made dichotomy between expert and non–expert in the use of bibliometric indicators. Thus, rather than reiterating accusations of indicator misuse by amateur- or layman-bibliometricians, we explore a much less frequently trodden path, namely how bibliometrics are used to defend and define notions of excellence in the evaluation of researchers across fields.

To put our findings into context, the following section sets out previous studies on researchers' uses and understanding towards bibliometrics, and introduce the concepts we will employ to help make sense of indicator uses in our empirical materials. We then consider the significance of assessment reports as data sources and lay–out the methods through which we analyzed the material. Our findings are then presented, starting with an overview of metric uses in the two studied fields, followed by a more detailed analysis of what we call 'the context of bibliometrics', which includes how different types of indicator expertise informed the uses of bibliometrics in these assessment reports. We conclude by discussing how a conceptualization of bibliometric indicators as judgment devices might further our understating of the concept of 'citizen bibliometrics' and what our proposals might entail for the future study of evaluative bibliometrics.

## 1.1 Researchers understanding of bibliometric indicators

Scientometricians have criticized the statistical properties of indicators such as the JIF and h-index (f.e. Van Leeuwen 2008; Larivière, Lozano, and Gingras 2014; Waltman and van Eck 2012), and similar criticisms has been directed towards journal ranking and rating lists in economics (Tourish and Willmott 2015). The importance of these criticisms cannot be underestimated, but our approach resembles more *verstehen*-type studies interested in how researchers make sense of indicators. Previous findings indicate that researchers' perceptions of citations and citation–based measures are ambivalent (Buela-Casal and Zych 2012; Hargens and Schuman 1990). For example respondents in a survey by Aksnes and Rip (2009) stated that they were knowledgeable about citations, but at the same they do not keep track of citations to their own work. Partly, these results could be the consequences of researchers perceiving that to take too much interest in one's own citations is frowned upon. Researchers also form so called 'folk theories' about citations; where, for example, some claim that being trendy is more important than quality or that the citation rate of the paper is dependent on the status of the author (Aksnes and Rip 2009). In a later study by Derrick and Gillespie (2013) more than sixty percent of the respondents said that they would include bibliometric indicators in an application if they perceived it as advantageous, while an equal percentage of respondents (60%) agreed with the statement that indicators encourage researchers to 'cheat' and 'game' the system.

Whilst acknowledging the importance of researchers' perceptions of bibliometrics, we also believe that these ambivalent attitudes may partly reflect the discrepancy between attitudes towards indicators and their actual use. Hence, this study contributes to an emerging literature studying use – rather than statistical properties or perceptions – of indicators. Taking such an approach Rushforth and de Rijcke (2015) for instance showed how biomedical researchers rely on the JIF when making decisions on co-authorship and publication venue. Our study build on these more direct approaches towards further our understanding of how researchers use and make sense of metrics by exploring how indicator uses varied according to disciplinary cultures.

## 1.2 Why analyse reports?

To our knowledge this is the first study that systematically analyses the use of bibliometric indicators in assessment reports that evaluate and rank candidates for academic positions. As is largely typical of evaluation in higher education and research, expert peer review is the method of choice for ensuring such outcomes. In this particular hiring context peer review is performed remotely (Gläser and Laudel 2005) by individuals provided with the same information and asked to judge according to pre–defined criteria. This is a traditional method of peer review in which peer judgment is an important input in decisions taken by some other agent (Bozeman 1993)[1], in this case appointment committees in Swedish Universities. Referee candidate reports are an important empirical resource through which to explore uses of bibliometric indicators as judgment devices in research evaluation, as such documents feature ubiquitously to evaluate and eventually rank careers within the Swedish 'academic market'. Moreover, these reports also tend to articulate disciplinary norms for how researchers are evaluated, and deliberations made in these documents have significant influence on individual careers. The study of these documents therefore offers unique insights into the uses of bibliometrics in situations where a lot is at stake, and where indicators are used both for making and justifying complex decisions.

## 1.3 Recruitment procedures in Swedish academia

Studying assessment reports for academic positions in Swedish academia has two distinct advantages. First, according to 'offentlighetsprincipen' (openness principle) all documentation on decisions made by state institutions in Sweden should by law be accessible to the public. Second, the procedure of external recruitment is fairly similar among institutions for higher education where external assessments, together with interviews and invited lectures by leading candidates, form the basis from which a formal hiring decision is made.

The tradition of using external appraisers, so called 'sakkunniga', has a long history in Swedish academia stretching back to the late 19$^{th}$ century. This system was introduced in order to preserve the legitimacy and independence of the university, and eventually these reports came to play a normative role (Nilsson 2009). Over time the importance of external assessment of research quality has lessened somewhat, as other merits such as teaching and administrative skills have been given more weight, yet skills in teaching are still usually overtrumped by research merits (Brommesson, Erlingsson, Karlsson Schaffer, Ödalen, and Fogelgren 2016).

The recruitment procedure in Swedish academia is designed to be impartial and merit–based in the sense that external reviewers are the ones assessing the candidates, yet there are many ways in which the recruiting department can influence the process. The broader politics and practices of academic recruitment is indeed a fascinating topic, which so far only briefly has been covered by literature on say academic job markets (cf. Musselin, 2009). However, in this study we zoom in on one specific part of this process: the assessment of research merits within external expert reports, with a special focus on the use of bibliometrics for assessing candidates in the external peer review stage of hiring and promoting.

## 2. Indicators and rankings as judgment devices

When external referees are assigned the task of providing a ranking of candidates they are called upon to provide order and reduce uncertainty. Their assignment is particularly difficult as the individuals being evaluated are unique, multifaceted, and complex. Each candidate has unique competencies which cannot be compared directly, the information provided by each applicant is at least partly distinctive, even if general criteria exists and there are some agreed-upon rules for how the assessment should be performed. Borrowing from economic sociology we find a parallel between evaluating these candidates and the valuing of unique goods, or what Lucien Karpik (2010) terms *singularities*. A singularity is a good that is unique and not readily compared to other products or services, such as a work of art, a novel, or a medical doctor. The difficulty of assessing singularities makes external support necessary in order to reach a decision. Customers, or in our case referees, rely upon external support in the form of *judgment devices* that facilitate and legitimate arguments and decisions. Judgment devices can, according to Karpik, be divided into five main types: *networks*, *appellations*, *cicerones*, *rankings* and *confluences*. We suggest that two of these, appellations and rankings, are particularly useful for understanding the role of bibliometrics when used in the context of evaluating researchers. Appellations are brands and titles that assign meaning and worth to a product or a group of products. In our case this could involve the brand of a journal (like *Nature*), but could also be a certification (e.g. journal indexed in *Web of Science*) or an origin (e.g. a journal/book published by *Oxford University Press*).

The effectiveness of appellations builds on shared conceptions regarding the identity and quality of a particularly label (Karpik 2010: 45-46). In cases where such agreement does not exist an option might be to instead make use of rankings. Rankings arrange singularities in a hierarchical list based on one or several criteria, and Karpik distinguishes between two different types of rankings: those that build on expert rankings and those that make use of buyers' choices. Expert rankings build on valuations made by domain specialists, and they could take the form of prizes annually awarded or public rankings of universities or hospitals. Buyers' rankings, on the other hand, are determined based upon the selling of particular products (e.g. 'top 10 lists' of most highly sold products). When consumers rely on judgment devices they agree to delegate decision–making to an external source, although they do not always understand how it works or have control over the device (Karpik 2010: 46). Eventually, as more trust is invested in these devices, the debate no longer comes to concern if a single device makes a fair valuation, so much as how different judgment devices stand up against each other. Here evaluation of goods is replaced by the evaluation of judgment devices (Karpik 2010).

Through our empirical findings, we outline the uses of bibliometric indicators as judgment devices in evaluations of candidates in the following ways: as references drawn on to substantiate claims about journal (article) quality in a candidate's CV; listing each candidate's rating on a specific measure without passing explicit comment, suggesting this practice as a formality of constructing an evaluation report for others to base decisions upon; and combining different judgment devices to cross–validate each other in support of a statement about one of more candidate.

## 3. Material and methods

In very general terms qualitative research is especially useful in expanding knowledge of exploratory topics where previous literature is either sparse or diffuse. We consider the uses of indicators in evaluation contexts to be one such instance. The analysis of bibliometric uses, which we will present in our findings section, emerged via an interpretive coding framework. To analyse the theme 'bibliometric uses' in the external assessment reports we followed an inductive content analysis approach, where our results are derived more from the data than a priori theoretical framework. The approach we took followed the *preparation*, *organizing*, and *reporting* phases set-out by Elo and Kyngäs (2008).

In the *preparation phase*, we gathered our empirical materials. The principle of openness allowed us to gather external assessment reports from four major universities in Sweden – University of Gothenburg, Lund University, Uppsala University, and Umeå University – which conduct research both in economics and biomedicine. The definition of economics was quite straightforward as the field was judged to be equivalent to the Swedish term 'nationalekonomi'. Biomedicine is a much more loosely defined term and we therefore included all specialist positions involving natural science applications in medicine. We collected material from a ten–year period starting in 2005 and ending in 2014 (table 1). We focused on external reports of applications, which were evaluated in competition, and cases with only one applicant were excluded. Joint statements by several examiners were treated as one report while independent reports pertaining to one particular case were treated as stand-alone documents. There were common structures to these reports, including a general introduction where the task at hand was to describe each candidate, which then led up to a ranking of applicants. Depending on the number of applicants and the ambition of the reviewer each of these reports range from a couple to over thirty pages.

**Table 1**. *Overview of studied material*

|  | Lund University | Umeå University | University of Gothenburg | Uppsala University | **Total** |
|---|---|---|---|---|---|
| *Biomedicine* | 46 reports | 3 reports | 22 reports | 61 reports | **132** reports |
| *Economics* | 17 reports | 4 reports | 27 reports | 8 reports | **56** reports |

External assessment reports for academic positions at state financed universities in Sweden are available to researchers without obtaining permission from the referees writing the reports or the candidates being assessed. While, both examiners and candidates probably are aware that colleagues or others interested in these processes may read these documents we still decided not to reveal the identity of either referees or candidates. Our decision rests on two premises: (1) neither the referees or the candidates had any opportunity to decide if they wanted these document to be part of a research project, and (2) we did not find that revealing the identity of candidates or examiners would improve the analysis. Consequently, all reports were coded based on year, field (biomedicine: bio, or economics: eco) and university (Lund University: LU, University of Gothenburg: GU, Uppsala University: UU, Umeå University: UMU). Although it was decided that both authors would be involved in the data organizing process, as only one is a native Swedish speaker, the other author was restricted to coding only those documents written in the English language (n =109). Quotes from Swedish language documents which are displayed in the findings section are based on the co-author's translation into English.

In the next phase of our inductive content analysis we sought to *organize* the collected data. This iterative process involved first careful readings and open-coding of the texts, followed-by re-reading and grouping emerging categories into more refined categories. Finally this phase involved 'abstraction', whereby sub-categories which fall under the 'main category' (i.e. bibliometric uses) are grouped and refined hierarchically under content-characteristic terms (e.g. JIF, rankings, H-index etc.). Although adopting a rather generous and open-ended definition of 'bibliometric use' in the early stages of coding, these subsequent steps allowed us to arrive at a more precise definition of our unit of analysis, restricting 'bibliometric uses' to instances where the JIF, h–index, journal rankings, or citations were actually employed with references to numbers in the reports. As such, the reporting phase of our analysis excludes instances of what we had considered 'bibliometric use' in earlier open-coding rounds, for example omitting categories like 'simple publication counts', or categories derived from general remarks about 'high impact journals' or 'top journals'. Although likely to be inferred indirectly from implicit knowledge of bibliometric measures or rankings, we decided these latter categories were rather too general to be helpful in illustrating distinct 'metric cultures' in biomedicine or economics.

In *reporting* our findings, we present quotes which are illustrative of the most frequently occurring sub-categories that emerged in the analysis process. Where it is deemed helpful, we provide simple numerical counts of the incidences of major and sub-categories of 'bibliometric uses' in the texts. These are meant to give an impression of how often uses of different metrics are made, not to make statistically generalizable claims about the incidences of bibliometrics in different evaluation cultures.

Based on our sample size and the strict definition of metric use we arrived at, we found their may be some risk in over-stating the presence of bibliometric indicators in the context of evaluating candidates for academic positions. For example, a simple count revealed 82 out of 188 of the studied reports made what we have come to define as explicit use of metrics. Nonetheless, given we are not aiming for statistical generalizability and the relevance of indicators in these remaining 82 texts, we suggest the study of bibliometric uses in concrete assessment contexts merits greater attention within the research evaluation literature. The value of our qualitative approach is in opening-up how metrics are drawn on to substantiate, question, and negotiate specific claims within a familiar yet underexplored research evaluation setting in different disciplinary domains. This enables us to draw (modest) conclusions and formulate further research questions about the role of bibliometric indicators as 'judgment devices' in research evaluation settings and to flesh out our own tentative insights in relation to the emerging concept of 'citizen bibliometrics'.

## 4. Findings

### 4.1 Using metrics in biomedicine and economics

In this section, we compare the uses of different metrics across the fields of biomedicine and economics. From those reports in which indicator usage was found we observed some similarities across disciplines, for example in reports where indicators were introduced without hesitation, and with an assumption that they directly reflect the scientific ability of the applicant:

> "Scientific skill has been judged based on scientific publications and citations to these publications registered in Scopus (www.Scopus.com) as well as h–index which highlights the quantitative influence of the author or scientific impact." (Bio GU 201–5, p. 1).[2]

> "A bibliometric analysis was carried out to assess the scientific production and even more importantly, the real scientific impact of each applicant." (Bio LU 2014–3, p. 5)

In the latter quote it is not even that metrics 'highlight' certain aspects, but it is said to represent *real* scientific impact, implying more qualitative descriptions may result in inaccurate assessment of applicants. The use of metrics is also motivated by indicators being 'unbiased' (Bio UU 2008–2, p.1; Bio UU 2012–11, p.1) and therefore providing fairer assessments of candidates. Although such strong sentiments advocating bibliometrics are rare we found that metrics were often presented in a neutral or positive tone across both disciplines.

Another explanation validating uses of metrics is the sheer volume of information that reviewers have to take into account, rather than say the uniqueness of each multi–dimensional object:

> Expert appraisals can be rather long tomes, and so I have attempted to utilize tables to compact the information provided by the candidates and available from other sources, such as the Web of Science. (Bio UU 2012–4, p. 1)

Resonating with Karpik's account of judgment devices, this quote suggests reviewers face problems with *excess*, both in the sense that there are several possible candidates and an abundance of information regarding these candidates. For the individual this may result in overload and habituation, a situation which might be solved either through the reduction of excess or through redefining excess (Abbott 2014). We suggest use of bibliometrics equates to a reactive strategy for reducing excess by "...hierarchizing and concentrating one's attention at the top end of the hierarchy." (Abbott 2014:18). In many cases reviewers point an abundance of quality candidates, rather than scarcity. In this situation ranking and rating devices offer a means of sorting between 'good' versus 'very good' candidates.

These arguments for using bibliometrics possibly point towards epistemic practices of evaluation in which expert judgments are often legitimated by way of mechanical, standardized, 'objective' indicators (c.f. Porter 1996). This resonates also with Lamont's (2009) observation that realist commitments toward objectivity in academic peer review often stem from epistemological traditions in which the evaluators are embedded as researchers (with peers deriving from social constructionist fields by contrast feeling more comfortable with the 'subjective' dimensions of peer review processes and skeptical towards bibliometrics). Although in a broad sense these observations appear tenable, as we will show, bibliometric use is highly context dependent and in several respects differs considerably between biomedicine and economics.

A little less than half of all assessments in biomedicine made explicit use of bibliometric indicators; 58 out of 132 reports (44 %). Explicit use here is defined as the giving of numbers – JIF, h–index or citation scores – in the actual text. Among the more frequent indicators used in biomedicine were the h–index (26 reports 20 %) and the JIF (23 reports 17 %), while straight, or adjusted citation counts were found in 38 reports (29 %).

The proportion of reports in economics that use bibliometric indicators or journal rankings in assessing applicants is almost exactly the same as in biomedicine 24 out 56 (43 %). Journal rankings are not a bibliometric indicator in a proper sense, but they are often in part derived from bibliometric indicators and play a similar role as judgment devices in these texts. We therefore included journal rankings and ratings in our study. Straight or adjusted citation counts were the most common indicator in economics (13 reports) with journal rankings also being frequently used (11 reports). JIF was given in 9 reports while h–index was used only in 3 cases. These figures are not intended to support claims for statistical generalizability, but provide nonetheless a useful précis of how frequently different indicators featured across our materials.

Comparing the two fields we find that referees in biomedicine tend to use JIF scores and the h–index, while journal rankings, which is not used at all in biomedicine, form a tradition in economics. The use of JIF in biomedicine could be interpreted as a form of ratings, or in Karpik's vocabulary *appellations*. This type of judgment device is most effective when evaluation criteria are well recognized and agreed upon, as is often the case with research fields characterized by high dependency on other researchers and low uncertainty regarding goals and procedures (Whitley 2000). Consequently, the form of rating, or 'branding', used in biomedicine is effective due to a general agreement on research priorities, goals of research and publication practices. Economics on the other hand distinguishes itself from other social sciences in having a low degree of task and strategic uncertainty, and it is also characterized by a rather high degree of mutual dependency; a type of organization that Whitley (2000) describes as a 'partitioned bureaucracy '. Distinctive for economics is the hierarchical structure of the discipline, which is upheld by a reputational élite through its influence over the training of new economists, the communications system, and access to resources (Coats 1993: 42). The voluminous production of rankings – of institutions, journals and scholars – in economics appears illustrative of what Fourcade, Ollion & Algan (2015) claim is a predilection for hierarchies within the field. The material under study here, in which no less than five different rankings are used, appears to resonate with these more general characteristics. The extensive use of journal rankings also suggest that the most relevant literature in economics is found in a distinctive and rather small set of key journals. Thus, the comparatively insular nature of literature in economics, which is confirmed by bibliometric studies of citation patterns (Fourcade, Ollion & Algan, 2015), is a further factor which might help to contextualize the popularity of journal rankings within the field. In sum our findings appear consistent with Whitley's account of both economics and biomedicine as fields featuring characteristics – low task uncertainty and high degree of mutual dependency – which would seem conducive for extensive use of bibliometric indicators in recruitment procedures. However, already at the outset we also identify differences across these fields in the type of bibliometric indicators used, and discipline specific characteristics. These differences will now be made further visible in a detailed description of indicators as judgment devices.

## 4. 2 Uses of indicators as judgment devices

### 4.2.1 The journal impact factor

In this section we show how indicators have come to assume different meanings and acquire different uses as *judgment devices* within our materials. The journal impact factor (JIF), built on an idea forwarded by Gross and Gross in 1927 (Gross and Gross 1927), later realized by Eugene Garfield and incorporated as a feature in the Science Citation Index (Garfield 1963), is one of the most popular and at the same time most criticized bibliometric indicators (Archambault and Larivière 2009). By calculating the average number of citations per article in a journal the JIF is said to give an indication of the 'impact' and relative standing of a periodical. In our materials we see how JIFs are used to establish orders and values among publications in the reports. A common practice is to attach JIFs in the text to support statements on journal quality. This is often done in a 'neutral' reporting type fashion with the JIF being given in a parenthesis after the name of the journal – almost like a reference in scholarly text to substantiate a statement, in this case the importance of a journal:

> " ...but it is a bit bothersome that many of the recent publications that
> XXXX has been principal investigator on are found in more narrow
> journals, as for example Scandinavian J Immunol. (Impact approx. 2.3)."
> (Bio GU 2012–2, p. 2) [3]
>
> His CV includes 20 peer reviewed publications in journals such as Physica
> D, Studies in Nonlinear Dynamics and Econometrics (Impact factor
> 0.593), European Journal of Health Economics, Applied Economics (2 st,
> Impact factor 0.473), (...) European Financial Management (Impact factor

0.717), Journal of Economics and Business, and Energy Economics
(Impact factor 1.557)." (Eco UU 2009-2, p. 2) [4]

Similar uses are found in several reports where JIF is given as supportive evidence of journals having a good reputation, or, as the illustration below makes clear (fig 1.), it can be presented in a table showcasing journals, their impact factor and the number of publications published in the same 'top journal' for each candidate.

| Journal | Impact factor | Number |
|---|---|---|
| Circulation | 10.94 | 3 |
| Eur heart J | 7.92 | 15 |
| JACC | 9.7 | 4 |
| Ann NY Ac Sci | 1.93 | 1 |
| Brit heart J | 3.7 | 5 |
| Pacing Clin Electro | 1.56 | 7 |

**Figure 1.** Table presenting journals, impact factors of these journals, and the number articles published by a candidate (Bio LU 2006-3, p. 4)

Another common use is to indicate a scoring interval (for example ranging from 4–7) of the JIF of journals where the papers of an applicant have been published (Bio GU 2006-1; Bio UU 2013-7). Such scales are taken to reflect not only the ability, but also the ambition of the researchers in question. Aiming, and subsequently succeeding in publishing in high impact journals signals a resourceful and successful applicant capable of overcoming peer review in journals with high rejection rates. The JIF appears to stand as an obvious shorthand for a journal's reputation and by association the standing of a candidate. However, in cases where the interval is broad – for example stretching from 0.5 to 26, or from ordinary to highest quality (Bio UU 2008–4) – such numbers carry little meaning and have to be supplemented by other judgment devices. A few referees take this a step further by aggregating JIFs and then coming up with an average or median of impact factors (Bio LU 2005-6; Bio UU 2012–4).

JIF scores can also be used for setting a standard or a benchmark. This is illustrated in a report claiming that most journals (in which the applicants have published) have 'a JIF over 4' (Bio UU 2012–7) or by counting the number of paper published in journals with an impact factor of 5 or better (Bio LU 2005-7). The magic number, for being regarded as of high quality publications in biomedicine, seems in many cases to be around 3 (Bio UU 211-3) or slightly higher:

> "Many of original papers appeared in excellent quality journals and
> nearly two third [sic] of them were in journals with impact factor greater
> than 3." (Bio LU 2014–4).

In economics, where citation frequencies are generally lower, we find statements claiming that journal impact factors of around 0.7-0.9 are normal for average field journals (Eco UU 2009-3), while others suggest that journals having impact factors over 0.5 are highly ranked (Eco GU 2010-1). Hence, what is to be considered as a high impact factor in the field of economics is not evident, and generally we find that impact factors in economics are accompanied by qualifying statements, which help the reader to evaluate the score which the judgment device produces.

In biomedicine especially, a JIF of a certain magnitude functions as a benchmark for what is to be considered a high quality journal, and eventually this 'stamp of quality' also serves as a device used for making judgments on the merits of individual researchers. The statement that specific journals have 'high impact', which explicitly or implicitly is derived from the long tradition of using the JIF in biomedicine, moves away from the context in which these numbers are produced, and becomes a 'fact' of its own. The JIF functions as an appellation, which assigns value to the journal, and eventually the impact factor becomes part of the 'brand' of the journal. Following this line of thought we would suggest that the influence of impact factors goes much further than their actual use, as they come to form a whole way of thinking and vocabulary for discussing quality.

### 4.2.2 The h-index

Whereas JIFs are often given in sections of the report where specific research contributions (in the form of journal articles) are discussed, the h–index is often given in a more general description of the applicant. Invented by Jorge Hirsch – a physicist and not a professional bibliometrican – the h–index is a very well-known attempt to come up with an indicator that reflects both the quality and the quantity of publications produced by an author. In short, if a scholar has an h–index of h it means that she has published h publications which has been cited at least h times (Hirsch 2005). We find that the h–index can be a discussion or integrated in the text, but to a lesser extent than with the JIF. Unlike the JIF, the h-index is often given as a standalone 'fact' about the applicant:

> XXXX publishes in good to very good journals including Plos Genetics, FASEB J, and Mol Biol Cell. H–factor=18. (Bio GU 2013–9, p. 3).

In other reports the h–index is given, together with other 'details' such as birth year, current position etc., to provide a background to the more narrative-based text (Eco UMU 2009–1, p. 8; Bio LU 2008-5). There are also examples where the h-index is introduced in the narrative using 'screen dumps' from Web of Science (Bio LU 2013-2). Here the h–index becomes closely connected to the person being evaluated; it helps to identify and characterize not only the scientific production, but also the applicant as such. Giving the h–index alongside other basic information also appears to heighten the importance of the measure as a necessary background fact, which is given before the actual narrative begins. Thus introducing metrics into the text as a 'mere formality' exhibits one way in which particular judgment devices can become taken–for–granted. The h–index score is not only commented upon, but is presumably expected to 'speak for itself' as an indicator of the candidate's relative standing in respect to individuals with whom she is being compared. There is thus an expectation that the persons reading the report will want to know this score and base their decisions in part upon it.

The h–index could be seen as an attempt to summarize a whole career in one single measure, and in some reports the h–index is represented as an almost magical number that can be used to characterize and grade a researcher.

The totalizing effects of using the h–index is illustrated by these numbers being hard to ignore once they are given, and there are several reports where they tend to play a decisive role. The most evident example is found in Bio UU 2014–1 where the h–index of each candidate corresponds almost perfectly with the assessment and rankings made. Upon closer inspection the final judgment made of 23 candidates for a professorship at Uppsala University (Bio UU 2014–1) aligns with the h-indexes of candidates (table 2).

**Table 2**. h–index and recommendation for professorship (Bio UU 2014–1)[4]

| *h–index* | *Recommendation by the referee* |
|---|---|
| 0–14 | Not qualified / not eligible |
| 15–20 | Borderline qualified |
| 21–25 | Qualified |
| 26–33 | Fully qualified |

Although certain judgment devices have become very important for evaluation in many fields, no one device completely dominates. This is evident in our materials, for example, the h–index quite often is combined with other indicators, such as straight citation counts or JIFs (Bio LU 2013-2; Bio UU 2012–4).

This section suggests forced decision–making situations such as these lend themselves to use of judgment devices, as reviewers must make recommendations amongst a range of 'commodities', all with unique multi–dimensional qualities ('singularities') (c.f. Karpik 2010). The authority and expertise of the reviewers is exercised in different ways across our materials. Whereas in some cases expertise was put forward via qualitative judgments of reviewers which are expected to carry weight in their own right, as we have made visible in this section and throughout the paper, expertise can also be displayed through using various bibliometric judgment devices. In the next section we consider further the dimensions of expertise demonstrated by uses of metric indicators in this evaluation context, and in doing so revisit previous characterizations of uses of bibliometrics as 'amateur', proposing instead the term 'citizen bibliometrics'.

## 4. 3 Citizen bibliometrics

In this paper we deliberately use the term 'citizen bibliometrics', first proposed in Wouters et al. (2015), as opposed to 'amateur bibliometrics' or 'non–professional use'. Rather than thinking of indicators as marking an 'outsourcing' of judgment and expertise to more mechanical 'objective' procedures, the term citizen bibliometrics allows us to concentrate on how indicators appear to be redefining what is meant by expert judgment in particular research fields. What we see in our material is indeed that some examiners are quite knowledgeable about bibliometric indicators and their shortcomings. For example, part of demonstrating expertise in these contexts comes from citing limitations of bibliometric indicators, as well as knowing which indicators to deploy in evaluating research outputs from their own field and knowing which not to.

Despite their wide acceptance, external assessors might still find it necessary to explicitly discuss the use of bibliometric measures. There are for example cases where the general assumptions, or folk theories, about JIFs appear as problematic when making judgments on applicants from different fields:

> Nuclear medicine journals do not have really high impact factors (not like
> e.g. Lancet and Nature having impact factors >20). The best journals focused
> on nuclear medicine most often has impact factors <10). (Bio LU 2014–6)

Interestingly, this justification is found in a report that does not make explicit use of JIFs, but the expert in this case, well aware of the tacit knowledge among readers of the report, wishes to make this difference explicit. Besides knowing the limitations of certain indicators, some examiners make judgments on which ones to use:

> I do not use citations as they are unreliable when citation windows are
> short and unevenly distributed. When it comes to journal impact the
> results are dependent on the measures used to a considerable degree.
> (Eco GU 2013–1, p. 8)[5]

As well as questioning the reliability of indicators, some also discuss the validity of indicators:

> Impact measures of this kind are inexact and should not, in our view, be
> relied on for a detailed ranking of research achievements (it could be
> described as 'a scale that can distinguish an elephant from a rabbit but
> not a horse from a cow'). However, the ability to publish in influential
> and selective journals is important in the Economics field and therefore
> such rankings provide useful information on the quality of research.
> (Eco GU 2012–2, p. 1)

In this case the reviewer hesitates when introducing metrics, and questions if metrics can be used to rank candidates, especially in cases where they have similar merits. Yet, the reviewer still finds them useful as they reflect an ability to publish in highly ranked journals – a skill that is valued in economics. The gist of the argument seems to be that the use of these indicators is justified because it is an accepted method for valuing research within the field, and not because the measures as such are very sophisticated. This sentiment is also evident in the following quote from an assessment report in economics:

> The ability to publish in influential and selective journals is important in the field of economics, which means that the ranking of journals influence the assessment of research quality. (Eco UMU 2012–2)[6]

In our material we find that the question of whether metrics should be used at all is supplemented by a discussion on how indicators are best employed for making judgments of candidates. There are indeed some examiners that are hesitant towards the use of bibliometric indicators, as in a case where a separation between 'assessing the candidates' and 'consulting' a database is emphasized:

> "After a first review of the applications I consulted 'ISI Web of Knowledge' for citations– and publication numbers. This has not changed my evaluation or my assessment of the candidates. I am aware that professor xxx has included these analysis (in extenso) in his evaluation, and therefore are they not included here." (Bio LU 2011-1, p. 1)[7]

Notably, this examiner not only desists the temptation of using metrics when preforming his evaluation, but also subtly criticizes this colleague for relying too much – 'in extenso' – on metrics.

Besides discussions on the applicability of indicators more generally we also see that more technical aspects are discussed such as the length of the citation window (Eco GU 2008–2, p. 1, Bio UU 2012–4), or the usefulness of specific databases:

> Apparently it takes time to make an impact in World of Knowledge [sic. Web of Science] and this limited information source is not useful for discriminating between applicants. An alternative, with a larger coverage, is Google Scholar and here we find rather large differences. (Eco LU 2010–3, p. 12)[8]

Comparisons are not only made between different bibliometric data sources, but also concerning specific indicators used. External examiners come to negotiate the results of bibliometric measurement by introducing other metrics that question established indicators; or in other words they question one judgment device by introducing another. For example one examiner touches upon the relation between the impact of a journal and that of an article published within the same journal:

> It might very well be that a highly cited article in a low ranked journal
> should be given a higher value than a rarely cited article in a highly
> ranked journal. (Eco GU 2008–4, p. 2)[9]

This argument does not dispense altogether with the notion that journal rankings have value, but importantly qualifies that the meaning of such indicators can and should be taken as simply one possible tool for informing merit. There are also examples when different indicators are juxtaposed to give a more complete overview of a candidate, as in this example where several numbers (published papers, authorship, max citations and h–index) are given:

> Of 44 published papers she is 1st author on 12 and senior author on 20.
> She has a surprisingly low citation rate, albeit with a high h–index (Max
> citation <60 in 2010, h–index = 17, Web of Science). (Bio UU 2012–11,
> p. 8)

Experts might also reflect on problems of data coverage or other crucial issues such as author ambiguity; the problem of tying publications to a specific author (Bio UU 2012–9).

Introducing time as a factor for contextualizing publication numbers and citation scores is another a way of negotiating results. Time in the sense of years as an active researcher is often introduced as an issue to consider, and in one assessment report in economics this is done by introducing all applicants with their names tightly followed by 'Years out' and 'citations': **"Years out: 14, Citations 3328"** (Eco GU 2014–2, p. 1.) This information is given before any descriptive text and the numbers are highlighted in bold by the referee in order to reflect their significance. Overall, it is common that indicators are contextualized or even adjusted for time, especially in cases where there are major differences in the professional age of applicants. Temporal aspects may also be considered when future impact is considered, and expectations may influence judgments:

> He has the lowest citation record among the three applicants (151
> citations and h–factor of 7, according to Harzing's PP ranking) but
> looking at the REPEC list of citations he has many recent citations, and
> my assessment is that his publication and citation profile will be very
> positive. (Eco LU 2011–8, p. 1).

This quote exhibits knowledge about the reliability of database sources within the field of economics (REPEC), combined with working assumptions about what constitutes an acceptable number of citations within a certain time window, which an advanced bibliometrician working outside of the field of economics would not be able to judge.

Finally, we also see several examples of examiners that bring in a range of indicators together with the purpose of comparing them with each other. This is usually done separately from the text in the form of tables. One of the most ambitious examples of this practice is found in Fig 2.

|  | Candidate A | Candidate B | Candidate C | Candidate D | Candidate E | Candidate F |
|---|---|---|---|---|---|---|
| No. of published papers/reviews | 44 / 5 | 26 / 1 | 5 / 0 | 34 / 1 | 9 / 3 | 47 / 3 |
| First / last author[1] | 6 / 11 | 7 / 8 | 3 / 0 | 10 / 1 | 4 / 0 | 18 / 10 |
| other senior author[2] |  |  |  | Author X (27/34 papers) | Author Y (6/9 papers) | Author Z (37/47 papers) |
| Selected papers[3]: | 10 | 10 | 4[6] | 17 (+1 poster) | 12 | 19 |
| median (range) i.p. | 2.48 (2.11-4.13) | 2.74 (1.33-31.20) | 3.34 (1.86-8.28) | 4.19[7] (1.52-35.53) | 9.68[8] (5.53-11.66) | 3.18 (2.52-9.68) |
| median (range) citation rate[4] | 9.5 (3-83) | 12 (4-81) | 22 (6 & 38) | 18.5 (5-58) | 26 (6-50)[8] | 22 (3-226) |
| Estimated h-index[5] | 16 | 10 | 2 | 14 | 10 | 20 |

**Fig. 2**: Comparison of candidates (columns) according to different indicators (rows) by one reviewer (anonymised) (Bio UU 2012–4, p. 5)

This table introduces the section 'scientific merits' and is accompanied by eight footnotes, which describe and compare these numbers. There are several clues pointing to an examiner that is accomplished in bibliometric analysis. Indicators are introduced and explained – and in the case of the h–index even referenced – and detailed considerations are made. For example, in footnote 7 it is stated that: "The number is lower (2.44) for the six papers where she is first author, but the citation rate, in many ways a more important measure, is not much different" (Bio UU 2012–4, p. 5). Here are two key qualifications made: citations are more important than impact factors and authorship order is important to consider. Moreover, it is interesting to note that the median (and range) of citations and impact factors are given rather than the average, a practice that seems advisable from a statistical perspective as citation scores generally are highly skewed. In this example several judgment devices are brought in and their strengths and weakness are scrutinized in an informed discussion.

Another means of demonstrating expert knowledge of indicators comes in referencing well–known critiques of mainstream indicators. Reviewers in several instances were aware of the major criticisms of the h–index, including its dependence on the academic age of the candidate (making it an indicator of age as much as one of impact). Several of the examiners were aware of this weakness, and some even came up with ways of solving this issue:

> It could be worthwhile to compare the bibliometric scores for the three
> strongest candidates. Their h–index are 9 (xxxx), 13 (yyyy) and 14 (zzzz).
> Their academic career is of different length which makes it interesting to
> study h–index divided in years after PhD–defence (minus parental leave):
> yyyy 0.68; zzzz 1.56 and xxxx 0.9. However, zzzz is not senior author of

> these publications which yyyy and xxxx is on some of the publications
> contributing to their h–index. (Bio UU 2010-2)[10]

The detailed and knowledgeable use of metrics shown in these examples points to the fact that many, but by no means a majority, of all examiners have considerable skills in handling, presenting, and contextualizing bibliometric data. We would therefore suggest that examiners in some of our documents emerge as experts in three roles: (1) as domain experts (2) experts on metrics and (3) experts on how metrics are used and valued within their field. Furthermore, our findings suggest that a move from evaluating publications (or products in Karpik's vocabulary) to evaluating judgment devices (bibliometric indicators) is evident in some reports. However, in most cases we find that judgments on actual content and significance of scholarly publications are combined with judgments on the indicators used to evaluate these publications. Expertise comes to be demonstrated through deploying a given judgment device that is used to evaluate in respective epistemic community, and through mediating between alternative judgement devices. As such, we envisage that the rather empty label of 'citizen bibliometrics' may help to evoke some of these expert dimensions which are seldom brought forward in discussions over 'amateur bibliometrics'.

## 5. Discussion

In this paper, we have proposed that a fruitful approach for understanding the attractiveness of bibliometric indicators is their ability to help form decisions in situations where the quality of a work, or an individual, is hard to assess. Accordingly, indicators can be understood as *judgment devices* in helping referees to reach a decision in a peer review context where there is a glut of suitable candidates. In competitive professional environments characterized by 'credential inflation' (Collins 1979), judgment devices play important roles in making distinctions between singularities (Karpik 2009). What, however, does the use of this concept and our empirical findings suggest about the character of 'citizen bibliometrics'? In this section we will outline the contribution of this paper, and expand on why the concept of citizen bibliometrics might be more appropriate as both an analytic and normative category than 'amateur bibliometrics'.

Our findings show that the use and development of bibliometric indicators is not restricted to professional bibliometricians, but these measures are to a considerable degree discussed, developed, and refined also by other groups. While acknowledging the great importance of knowledge about how indicators are constructed we want to emphasize that 'Citizen bibliometrics' should not be considered as simple use or misuse of indicators developed by bibliometric experts. On the contrary we find the efforts of developing more advanced indicators – over one hundred different indicators of impact has so far been developed by the bibliometric community (Wildgaard, Schneider, and Larsen 2014) – are largely ignored when used in the context of assessing individuals. Simple and well–established indicators, like the JIF and the h–index, are preferred. The reason is probably not only that these indicators are readily available and relatively easy to calculate, but also that they are well established form of evaluation with some disciplinary communities. Moreover, our study shows that domain specialists quite often possess a considerable knowledge about these measures and they are aware of limitations and field differences when using indicators. In contrast with bibliometric experts they also have knowledge about which indicators are valued and recognized within their own field. Clearly, these examiners are not amateurs in using bibliometrics – in fact they are actually paid to conduct these assessments – meaning clear-cut distinctions between expert or amateur, professional or non–professional use is difficult to make based on our material.

Often qualification statements were provided, demonstrating reflexivity about the deployment of indicators. Such qualifying arguments also imply the belief that others do indeed use these systems as de–contextualized representations of excellence or quality. Reviewers here justify the uses of indicators and their own expertise in using them *by citing their limitations*, a rhetorical strategy which is often employed also by experts of advanced bibliometrics. Again this suggests common distinctions between expert and amateur bibliometrics may be too sharply drawn, at least insofar as they do not account for intermediary modes of expertise we have reported in this paper. It is we believe these 'shades of grey' that populate evaluative bibliometric practices, and therefore which require further attention if the uses of bibliometric indicators in research evaluation are to be better understood and theorized. To do so we suggest that studying the context of indicator use is crucial.

Our analysis has led us to suggest that comparison of indicators as judgment devices across disciplines is productive and even crucial for understanding the influence of bibliometric measures as technologies of evaluation. The giving of h–index or JIF as proxies for impact will mean something different in biomedicine compared to economics, and these numbers are treated quite differently depending on the discipline in question. While bibliometric indicators are used across both disciplines we find that indicators serving as ratings (f.e. the JIF), or what Karpik (2010) calls appellations is an important part of evaluation in biomedicine, while journal rankings are more popular in the field of economics. In turn we have argued that differences in the social and intellectual structure of these fields are of great importance when analyzing evaluation procedures. For example the organization of economics facilitates a further use of journal rankings, which then accentuates and supports arguments about a preoccupation with reputational hierarchies. The importance of the journal impact factor in biomedicine, which in many way is firmly integrated also in the production of knowledge (Rushforth & de Rijcke, 2015), relies on rather standardized evaluation procedures, which in turn are dependent on widespread agreement on methods and priorities in the field. Overall, the organization of biomedicine and economics, as well as the publication practices of these fields allows for rather extensive, although not always warranted, use of bibliometric indicators. The idea that disciplinary differences are important for understanding the consequences of bibliometric evaluation is well established, but we suggest that combining a contextualized understanding of bibliometric indicators as judgment devices with a framework for characterizing the organization of specific fields opens-up greater potential for more detailed analyzes of bibliometric use. Such an approach also destabilizes current distinctions between expert– and amateur bibliometrics at the same time as it questions the traditional juxtaposition of (pure) peer review and mechanized indicator use. In fact these types of evaluation intermingle in our material and domain expertise and bibliometric expertise cannot always be separated.

In our case we deem that disciplinarity is one important context of bibliometric use, but national differences and institutional settings might be other important factors to consider for future research on the use of metrics. Furthermore, while our focus has been on the use of metrics we suggest that evaluation reports for academic positions – a rich but understudied material – may bring insights to how careers are assessed and how authorship is valued. Moreover, an analysis of the applications that are assessed (the documents provided by candidates for academic positions), not only the evaluation reports as in this case, might give further insight to the use of metrics in these settings. Efforts, like Nilsson (2009), to study these documents over time might also bring knowledge on how specific evaluation repertoires have evolved. For example the establishment of online platforms that claim to measure 'impact' might result in indicators, such as the *'ResearchGate* score', becoming further integrated to the evaluative practices of academic fields. In our material *ResearchGate* is mentioned once (Bio UU 2013-2), but due to the popularity of *ResearchGate* and similar platforms use of such measures is likely to increase. Lastly, while biomedicine and economics are both large and influential fields where extensive use of bibliometric indicators has already been reported in the literature, we also see a need for extending our approach to fields where the use, and possibly the knowledge about bibliometrics, is less widespread. For example, 'citizen bibliometrics' in the humanities would, apart from being less common, probably be manifested in rather different ways.

What use then might comparative studies of indicators as judgment devices across different fields be in understanding ongoing debates on evaluative bibliometrics? One aspect highlighted in this paper is that it can help to expand assumptions about what constitutes expertise when bibliometrics start to become embedded into evaluation contexts. One logical development in any context where judgment devices are used for reaching decisions is that evaluators start to assess, and compare the devices used to make judgments and not the entity being evaluated in the first place. In short they become experts in judgment devices as much as experts on the specific topic at hand. This is one component of citizen bibliometrics we would like to put forward. The expertise of such reviewers is not necessarily simply in their research specialty (as implied by notions of 'informed peer review'), but in mediating between their own epistemic cultures of evaluation and knowledge production on the one hand and the affordances and limitations of specific bibliometric indicators on the other. Thus, it is knowing *how* and *when* to deploy indicators which should be considered the marker of expertise in such evaluative contexts. We suggest the term 'citizen bibliometrics' as a more inclusive and generous means of conceptualizing evaluative expertise than 'amateur bibliometrics'. Moreover, the term 'citizen bibliometrics' also connotes an ethical and caring dimension, where the duty is to apply indicators in manners which do not damage the community which the evaluator serve. Therefore producing citizen bibliometricians that are both literate in uses and limitations of bibliometric indicators and knowledgeable about their application in their own specific epistemic domain is an important political challenge facing the future of evaluative bibliometrics.

**Notes**

[1] Other examples of traditional peer review include journal peer review in which the executive decision on accepting or rejecting a manuscript is taken by an editor based on information produced by remote referees.

[2] Original (Swedish): " Den vetenskapliga skickligheten har bedömts baserat på vetenskapliga publikationer och publikationernas citeringsgrad dokumenterad i Scopus (www.Scopus.com) samt h–index som belyser författarens kvantitativa påverkansgrad eller vetenskaplig genomslagskraft."

[3] Original (Swedish): " Många publikationer är i goda tidskrifter, men lite bekymmersamt är att många av de som xxxx är huvudansvarig för på senare tid återfinns i lite smalare journaler som t.ex. Scandinavian J Immunol. (Impact ca 2.3)"

[4] Original (Swedish) " Hans CV inkluderar 20 publikationer i tidskrifter med referee förfarande såsom Physica D, Studies in Nonlinear Dynamics and Econometrics (Impact factor 0.593), European Journal of Health Economics, Applied Economics (2 st, Impact factor 0.473) (...) European Financial Management (Impact factor 0.717), Journal of Economics and Business, och Energy Economics (Impact factor 1.557)."

[5] There is only one case out of 23 assessments where an assessment diverged from this pattern – a researcher with an h–index of 16 which was deemed as 'not qualified.

[6] Original (Swedish): "Jag använder inte citeringar eftersom de inte är pålitliga när citeringsfönstren är korta och ojämnt fördelade. När det gäller tidskriftsimpact beror resultaten i ganska stor utsträckning på vilket mått som används."

[7] Original (Swedish): "Förmågan att publicera i inflytelserika och selektiva tidskrifter är viktig i nationalekonomifältet, vilket innebar att rankningen av tidskrifter påverkar bedömningen av forskningens kvalitet."

[8] Original (Danish): ""Efter den første gennemgang af ansøgningerne har jeg konsulteret "lSI Web of Knowledge" citations– og publiceringstal. Det har ikke ændret min vurdering eller rangordning af ansøgerne, men blot understreget betimeligheden af den primære vurdering. Då jeg er vidende om, at professor xxx inkluderer de nævnte analyser (in extenso) i sin vurdering, er de ikke medtaget her."

[9] Original (Swedish): "Uppenbarligen tar det tid att göra avtryck i World of Knowledge

och för att diskriminera mellan de sökande blir den lilla informationskällan inte så användbar. Ett alternativ som har bredare träffyta är därför Google Scholar och här framgår relativt stora skillnader."

[10] Original (Swedish): "Det kan har mycket väl vara fallet att en ofta citerad artikel i en lågt rankad tidskrift ska varderas högre än en sällan citerad artikel i en högt rankad tidskrift."

[11] Original (Swedish): Det kan vara värdefullt att först jämföra en bibliometrisk parameter för de 3 starkaste kandidaterna. De tre starkaste kandidaternas h–index är 9 (xxx), 13 (xxx) och 14 (xxx). Forskarkarriärernas längd efter disputation varierar mellan de tre kandidaterna, varför det är intressant att studera h–index dividerat med antal år (minus föräldraledighet) efter disputation: xxx 0.68; xxx 1.56; xxx 0.9. xxx är dock inte senior författare på de aktuella uppsatserna, vilket xxx och xxx är på en del av de uppsatser som bidrar till deras h–index.